\documentclass[reprint,amsmath,amssymb,aps,superscriptaddress]{revtex4-2}

\usepackage{graphicx}
\usepackage{dcolumn}
\usepackage{bm}
\usepackage{tikz}
\usetikzlibrary{quantikz}
\usepackage{braket}
\usepackage{multirow}
\usepackage{amsmath}
\usepackage{float}
\usepackage{hyperref}
\hypersetup{pdfborderstyle={/S/U/W 1},pdfborder=0 0 1}
\usepackage{placeins}
\usepackage{svg}
\usepackage{verbatim}

\newcommand{\ColorComment}[3]{%
	{\colorbox{#1}{\color{white}   \textsf{\textbf{#2}}} \textcolor{#1}{#3}}}

\definecolor{mscolor}{rgb}{0,0.5,0.5}\newcommand{\ms}[1]{\ColorComment{mscolor}{ms}{#1}}

\definecolor{cpcolor}{rgb}{0.5,0.5,0}

\definecolor{tgcolor}{rgb}{0.5,0,0.5}
\definecolor{nwcolor}{rgb}{0,0.5,0}

\newcommand{\InfleqtionM}{Infleqtion, Inc., Madison, WI, 53703, USA}
\newcommand{\UWM}{Department of Physics, University of Wisconsin-Madison, 1150 University Avenue, Madison, WI, USA}
\newcommand{\WHS}{West High School, 30 Ash St.
Madison, WI, 53726, USA}

\begin{document}

\title{Benchmarking a Neutral-Atom Quantum Computer}

\author{N. Wagner}
\affiliation{\WHS}
\affiliation{\UWM}
\author{C. Poole}
\affiliation{\UWM}
\author{T. M. Graham}
\affiliation{\UWM}
\author{M. Saffman}
\affiliation{\UWM}
\affiliation{\InfleqtionM}

\date{\today}

\begin{abstract}
In this study, we simulated the algorithmic performance of a small neutral atom quantum computer and compared its performance when operating with all-to-all versus nearest-neighbor connectivity. This comparison was made using a suite of algorithmic benchmarks developed by the Quantum Economic Development Consortium. Circuits were simulated with a noise model consistent with experimental data from Nature {\bf 604}, 457 (2022). We find that all-to-all connectivity improves simulated circuit fidelity by $10\%-15\%$, compared to nearest-neighbor connectivity.
\end{abstract}

\maketitle

\section{\label{sec:level1} Introduction}

Many candidate platforms for the physical realization of quantum computing hardware have emerged in recent years, including neutral atoms, 
trapped ions,
superconductors, 
quantum dots,
 and photonics\cite{Ladd2010,Bergou2021}. All architectures suffer from various types of errors that functionally limit the number of qubits (circuit width) and gates (circuit depth) that can be executed in a circuit while maintaining high fidelity results. 

Several different benchmarks have been proposed to compare the performance of quantum computers\cite{Cross2019,Wack2021,Allouche2021,Tomesh2022}. We have chosen to use a benchmark developed by the Quantum Economic Development Consortium (QED-C benchmark), which uses the  fidelities of several quantum circuits, such as the Quantum Fourier Transform and Grover's algorithm, to quantify the performance of a computer \cite{Lubinski2021}. 

 In this work, we present a noise model developed for a neutral atom quantum processor and benchmark simulated circuit execution. We constructed the noise model through a combination of diagnostic measurements and fitting simulated bitstring probabilities to measurements of circuits previously run on the computer. The quantum processor uses neutral atom qubits  on a 2D square grid with gates implemented using lasers and microwave fields \cite{Graham2022}. Because some circuits in the QED-C benchmark require more qubits than were used in the quantum processor, we assumed the errors are  uniform on each site and extended the simulator to accommodate high-width circuits provided in the benchmark. We ran simulations with both all-to-all and nearest-neighbor connectivity to accommodate the range of potential future device connectivity capabilities. We found significant improvement of circuit fidelities for the all-to-all connectivity compared to nearest-neighbor connectivity. 

The article is organized as follows: in Sec. \ref{sec:Methods}, we  discuss the methods by which we constructed the noise model for the simulation and give an overview of the QED-C benchmark. In Sec. \ref{sec:Results}, we present the results of the simulated benchmark circuits. We conclude in Sec. \ref{sec:conc} with a  discussion of  the significance of the results.

\section{Methods}
\label{sec:Methods}

\begin{table*}[!t]
    \centering
    \begin{ruledtabular}
    \begin{tabular}{cllc}
        Gate/Process & Noise & Error & Avg. Gate Fidelity \\\hline
        Global ${\sf R}^G_{\phi}$ (per $\pi$ pulse) & Depolarization  & $1.8\times 10^{-6}$ & $0.9995$\\ 
        \hline
        \multirow{4}{*}{Local ${\sf R}_z$ (per $\pi$ pulse)} & Phaseflip & $3.2\times10^{-4}$ 
        & \multirow{4}{*}{0.995}\\ &Loss to Dark State  & $1.9 \times 10^{-4}$ &\\
        & Loss to Bright State  & $2.7 \times 10^{-4}$ &\\ & Decay  & $2.0\times 10^{-8}$ &\\
        \hline
        \multirow{5}{*}{$\sf C_{\sf Z}$} & Phaseflip & $3.3\times10^{-2}$ & \multirow{5}{*}{0.954}\\ & Loss to Dark State & $1.8 \times 10^{-2}$ &\\ & Loss to Bright State & $2.9\times10^{-2}$ &\\ &Decay & $2.1\times 10^{-5}$ \\
        & Phaseshift & $-2.0 \times 10^{-3} \textnormal{ rad.}$ &\\
        \hline
        \multirow{2}{*}{SPAM} & Preparation & $5.2 \times 10^{-3}$ & \multirow{2}{*}{NA}\\
        &Measurement&  $5.3\times 10^{-3}$&\\
        \hline
        \multirow{3}{*}{Decoherence} & $T_1=10~\rm s$ &  &
        \multirow{3}{*}{NA}\\& $T_2^\ast=3.5~\rm ms$ & NA & \\
        & $P_{\ket{0}}=0.42$ at $t=\infty$ &  &
    \end{tabular}
    \end{ruledtabular} \\
    \caption{ \label{tab:table1} List of the types of error used in the noise model, along with the probabilities of each being applied. The noise channels and rates are organized by gate. Note that in some cases multiple gates share the same noise channel at different rates. The indicated gate fidelities in the last column are the entanglement fidelities obtained directly from the density matrix of a single application of a gate on a random pure state (not Hellinger distance), and are SPAM-corrected.   For phase-dependent channels, the error/avg. gate fidelity is per $\pi$ pulse. For additional details, see Appendix \ref{ssec:noisemodel}. 
    }
\end{table*}

\subsection{Hardware}

The neutral atom quantum processor we studied consists of a two-dimensional array of cesium atoms where qubits are encoded in the hyperfine clock states $\ket{0}=\ket{6s_{1/2},f=3,m=0},\ket{1}=\ket{6s_{1/2},f=4,m=0}$. Single qubit ${\sf R}_z$ gates were applied with a local Stark shift provided by focused 459-nm light that was blue-detuned from the $\ket{6s_{1/2},f=4} \rightarrow \ket{7p_{1/2}, f=4}$ transition. Controlled-Z $\sf (C_Z)$ gates were implemented using two-photon excitation of Rydberg states\cite{Levine2019}. Global rotations about an arbitrary axis in the $x-y$ plane of the Bloch sphere were performed with resonant microwave fields. The global rotations are written as ${\sf R}^G_{\phi}(\theta)$, where $\phi$ is the angle between the rotation axis and the $x$-axis, and $\theta$ is the rotation angle \cite{Graham2022}. These three gates provide a universal gate set from which any quantum circuit can be constructed. For example, local ${\sf R}_{\phi}(\theta)$ rotations may be implemented by combining global microwave rotations and local Stark shifts using the identity
\begin{equation}
   {\sf R}_\phi(\theta)={\sf R}_{\phi+\frac{\pi}{2}}^G \left(\frac{\pi}{2}\right){\sf R}_{z}(\theta){\sf R}_{\phi+\frac{\pi}{2}}^G\left(-\frac{\pi}{2}\right).
\end{equation}
This provides a rotation to the Stark shifted site, but the other atoms of the array see no net rotation since 
\begin{equation}
   {\sf I}={\sf R}_{\phi+\frac{\pi}{2}}^G \left(\frac{\pi}{2}\right){\sf R}_{\phi+\frac{\pi}{2}}^G\left(-\frac{\pi}{2}\right).
\end{equation}

The three gate types have different corresponding noise channels, as listed in Table \ref{tab:table1}. Independent experimental measurements determined $T_1$ and $T_2$ \cite{Graham2022}. We estimated the remaining noise parameters by fitting simulated bitstring populations against experimental populations of three 4-qubit QAOA-MaxCut circuits presented in the manuscript. The quality of the fit is determined by the classical fidelity between the two probability distributions, defined as
\begin{equation} 
F_{\rm s}(P_{\rm ideal},P_{\rm out}) = \left( \sum_x\sqrt{P_{\rm ideal}(x)P_{\rm out}(x)}\right)^2,
\label{eq.Fs}
\end{equation}
where $x$ is summed over all possible bitstrings, $P_{\rm ideal}$ is the ideal output, and $P_{\rm out}$ is the noisy output. This measure is then normalized to make the fidelity between the output state and the maximally mixed distribution 0.  The resulting normalized fidelity is defined by
\begin{equation}
    F_{\rm n}(P_{\rm ideal},P_{\rm out})=\frac{F_{\rm s}(P_{\rm ideal},P_{\rm out})-F_{\rm s}(P_{\rm ideal},P_{\rm uni})}{1-F_{\rm s}(P_{\rm ideal},P_{\rm uni})},
    \label{eq.hellinger}
\end{equation}
where the state $P_{\rm uni}$ is the maximally mixed multi-qubit state which has each diagonal element equal to $1/2^n$, with $n$ the number of qubits, and all off-diagonal entries equal to zero. This normalized fidelity can be negative in certain cases, so the final reported fidelity is defined as
\begin{equation}
\label{eq.classical_fidelity}
F(P_{\rm ideal},P_{\rm out})=\max\left(F_{\rm n}(P_{\rm ideal},P_{\rm out}),0.0\right).
\end{equation}
To quantify fidelities of the quantum gates $F_G$, we use the  definition 
\begin{equation} 
F_G(\rho,\sigma)=\left[ \textnormal{Tr}\left(\sqrt{\sqrt{\rho}\sigma\sqrt{\rho}}\right)\right]^2,
\end{equation}
where $\rho$ and $\sigma$ are density matrices of the ideal and simulated states. This measure takes both phase and population errors into account. The average fidelity of a quantum gate is calculated by averaging fidelity between ideal and simulated outputs of a set of Haar-random input states. The gate fidelities are reduced by various sources of error present in the physical implementation of the quantum gates.  All error channels are described further in Appendix \ref{ssec:noisemodel}.

\begin{figure}[!t]
    \includegraphics[width=9cm]{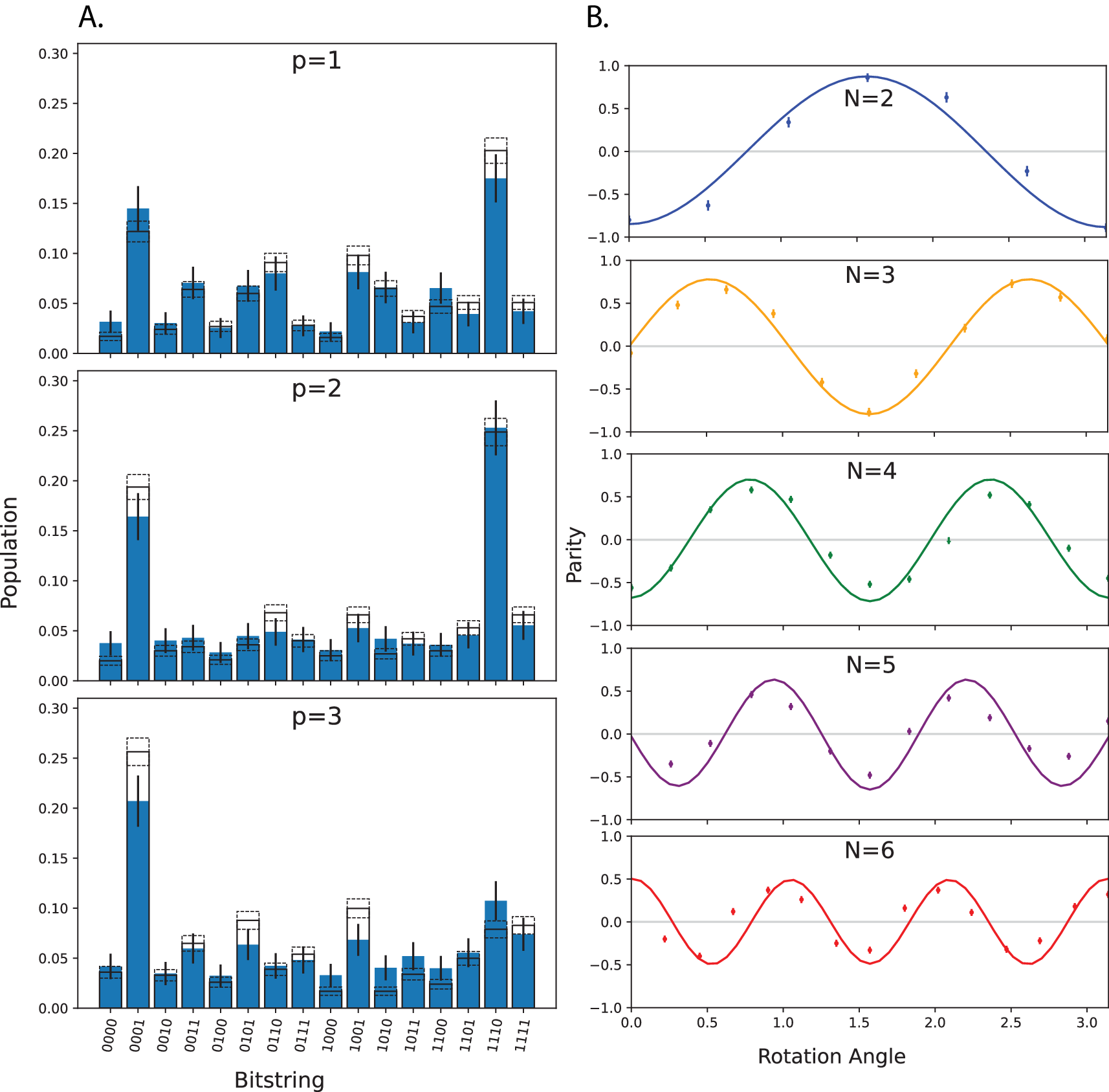}
    \caption{a) Simulation of 4-qubit MAXCUT algorithms from \cite{Graham2022} with $p=1$, $p=2$ and $p=3$. The unfilled bars represent experimental data and the blue fill represents simulated data. The lines represent error in the simulated data, and the dashed boxes represent error in the experimental data. b) Parity oscillation curves of 2-6 qubit GHZ states. The data points represent experimental data 
    , and the solid-colored curves represent simulated data.
    }
    \label{fig:ghz}
\end{figure}

The two-qubit $\sf C_Z$ gate uses Rydberg blockade to entangle qubits by applying a phase shift conditioned on the state of the qubits. During this gate, both qubits are partially excited to a high-energy Rydberg state outside of the computational basis. An ideal $\sf C_Z$ operation maps the Rydberg state populations back to the computational basis at the end of the gate. In practice, the population transfer is imperfect, and some qubits are either lost from the trap or to states outside of the computational basis. For simplicity, we treat leakage and atom loss events as either populating a dark lost state  $\ket{l}_0$ (for atoms lost from the array or leaked into $f=3, m \neq 0$ states that appear dark during the state-selective readout), or a bright lost state $\ket{l}_1$ (for atoms leaked into $f=4, m \neq 0$ states that appear bright during the state-selective readout), which does not participate in subsequent gate operations. To track the effects of these loss states in addition to the computational basis states, we simulate the computer as a ququart system rather than a qubit system. The $\sf C_Z$ gate is the most error-prone operation, with a simulated average gate fidelity of $\mathcal F\simeq  0.95$. The gate introduces a phaseflip channel, and a probability of qubit loss to a bright state $\ket{l}_1$ or dark state $\ket{l}_0$ from state $\ket{1}$. 
It also introduces a qubit decay channel, which transfers qubits in the $\ket{1}$ state to qubits in the $\ket{0}$ state.

\begin{figure*}[!t]
    \centering
    \makebox[\textwidth][c]{\includegraphics[width=16.cm]{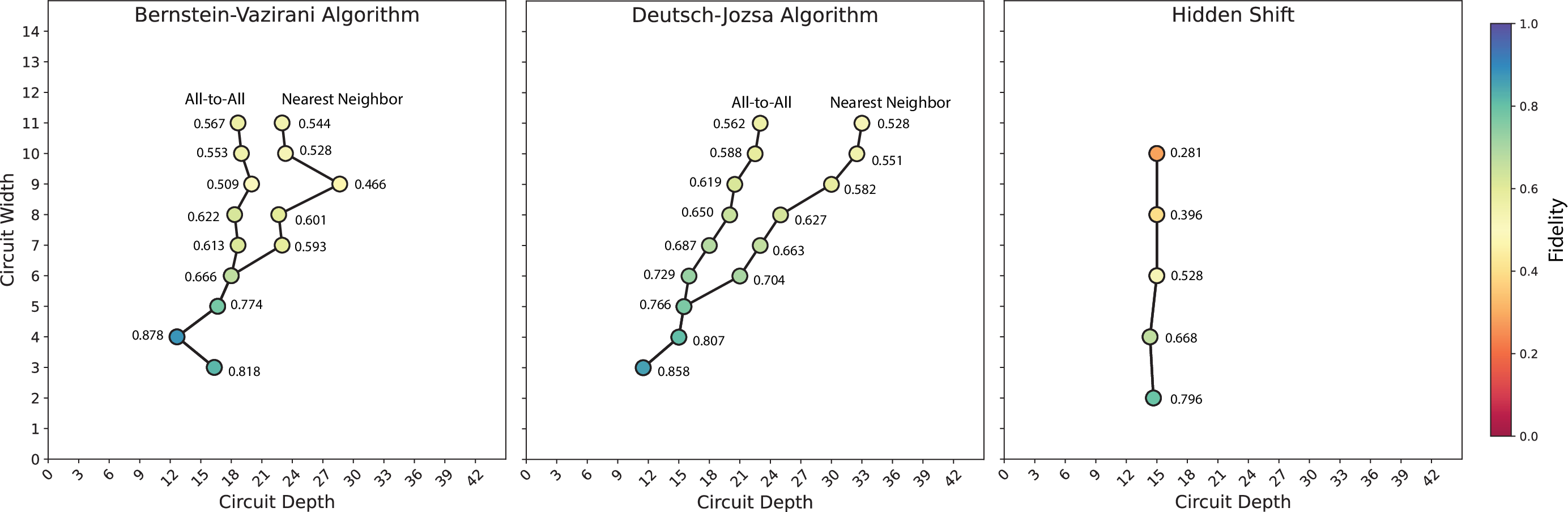}}
    \caption{Simulated results for the Deutsch-Jozsa, Bernstein-Vazirani, and Hidden Shift benchmarks for all-to-all and nearest-neighbor connectivity. For hidden shift  there was negligible difference between all-to-all and nearest-neighbor circuits.  The color bar shows the average fidelity of the result. }
    \label{fig:oracle}
\end{figure*}

The microwave and local ${\sf R}_z$ gates had the highest simulated fidelities with
$\mathcal F_{\sf R_\phi}\simeq 0.9995$ and $\mathcal F_{\sf R_Z}\simeq 0.995$. 
Microwave gate errors were modeled by following a perfect rotation with a probabilistic depolarization error. Local ${\sf R}_z$ gate errors were modeled by following a perfect rotation by a phase-flip, decay, and loss channels to both dark and bright lost states.
The $T_1$ decoherence time is $10^6$ times longer than the average gate execution time, making it negligible. Both the $\ket{0}$  and $\ket{1}$  states can scatter  into each other (primarily due to Raman scattering from the trap light), with a measured equilibrium population of $P_{\ket{0}}=0.42$ in the $\ket{0}$ state. $T_2$ decoherence, which is $\sim 100$ times greater than gate execution times, is more significant, and significantly degrades the performance of longer circuits.

State preparation and measurement (SPAM) errors were simulated to occur with about $0.5\%$ probability in both preparation and measurement. Preparation error acts as a global bit-flip channel proceeding the circuit, while measurement error randomly flips the bits of the output states before they are measured. 

The average classical fidelity (see Eq. \ref{eq.classical_fidelity}) between experimental and simulated populations for the three circuits was $98.6\%$, with a minimum of $97.5\%$ in the $p=3$ MAXCUT circuit. The average Bell state fidelity with these parameters is $91.3\%$, which agrees with experimental results in \cite{Graham2022}. Note that in the circuit producing a Bell state in \cite{Graham2022} a $\sf C_Z$ operation is performed on a state particularly sensitive to noise introduced by the $\sf C_Z$ gate. On average, simulated GHZ circuits overestimated the fidelity of the GHZ state by $8.0\%$.

These discrepancies are likely due to the fact that noise parameters vary between sites and change over time in experiment, while we assume them to be constant in the model. Additionally, some noisy channels, particularly channels with relatively low error, were simplified to improve the simulator's overall complexity. In particular, the choice of depolarization for microwave rotations does not fully describe the behavior of the gate. Despite these simplifications, we see close correspondence between the simulated and measured circuit results.

\subsection{Benchmark}

We used the QED-C benchmark because it includes a variety of circuits and is thus an effective measure of the computer's practicality. Unlike quantum volume, which only uses square circuits\cite{Cross2019}, the QED-C benchmark uses both circuits of high depth and low width and circuits with low depth and high width. As such, we can independently test the limits of the processor with respect to depth and width. Other benchmarks, such as the benchmark utilizing mirror circuits described in \cite{Wright2019}, also sample more circuits at lower circuit widths. However, such benchmarks use impractical circuits that are randomly generated. It has been shown that the predictions given by randomized benchmarks, such as these, are not a good predictor of the fidelity of practical circuits \cite{Lubinski2021}. 

The QED-C benchmark uses the fidelity results of common quantum computing circuits to establish a means to compare quantum computers. The circuits used for benchmarking are divided into three broad categories:
shallow oracle-based algorithms, quantum subroutines, and circuits applicable to real-world problems. Each circuit is run one or more times at different widths and with random inputs. The fidelities from multiple runs are averaged and plotted based on the circuit width and the average transpiled circuit depth. Note that the QED-C benchmark uses two methods to benchmark the performance of the quantum Fourier Transform (QFT), which we will refer to as Method 1 and Method 2. Method 1 encodes a qubit into the Fourier basis using a QFT, modifies it, and decodes it with an inverse QFT. Method 2 prepares a qubit in the Fourier basis using a global Hadamard gate and ${\sf R}_z$, then decodes it with an inverse QFT. In effect, Method 1 runs two consecutive QFTs, while Method 2 only runs one.

Since the computer did not have the capability to run some of the high-width circuits in \cite{Lubinski2021}, we sacrificed the accuracy of site-specific error parameters in the noise model for averages uniformly applied to all qubits. As shown in Fig. \ref{fig:ghz}, the simulation still accurately described experimental results, with the additional benefit that it was able to simulate a larger quantum computer.

We simulated the computer's performance in two cases; a best case, where connectivity is all-to-all, and a worst case, where connectivity is limited to nearest neighbors on a two-dimensional square grid. 
The long range nature of the Rydberg interaction makes it possible to consider beyond nearest-neighbor connectivity for moderate circuit widths. 
In this simulation, gate errors were assumed to be the same for both cases and independent of the distance between the two qubits. We simulated circuits with a width of up to 11 qubits. 

\section{Benchmark Results}
\label{sec:Results}
\begin{figure}
    \centering
     \includegraphics[width=8.5cm]{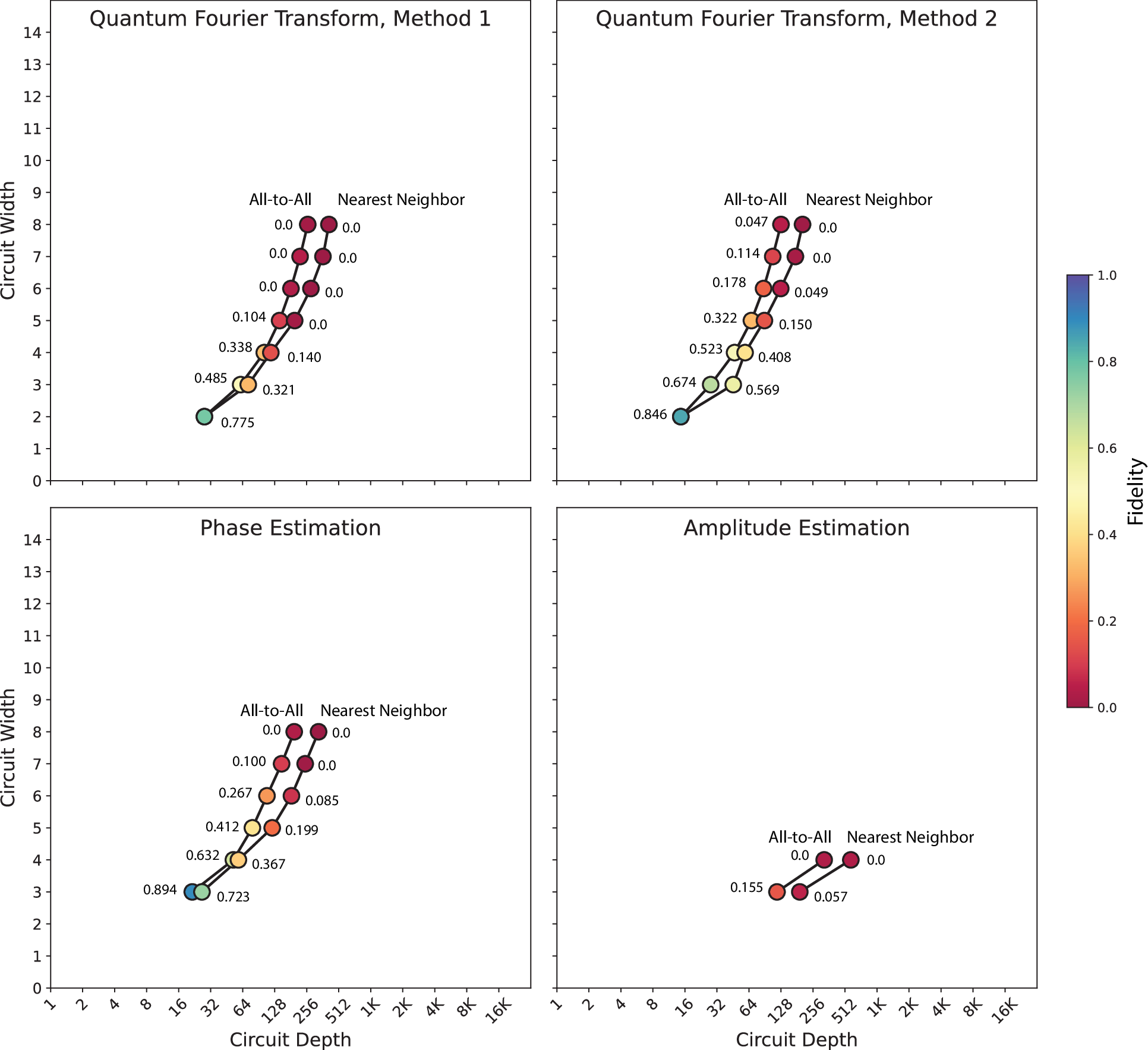}
    \caption{Simulation results for the phase estimation, amplitude estimation, and both Quantum Fourier Transform benchmarks.}
    \label{fig:subroutines}
\end{figure}

As shown in Fig. \ref{fig:oracle}, the computer simulation performed best with the shallower, oracle-based algorithms present in the QED-C benchmark (e.g. the Bernstein-Vazirani and Deutsch-Jozsa algorithms). With only nearest-neighbor connectivity, it ran the Bernstein-Vazirani and Deutsch-Jozsa circuits with up to 5 qubits with $>70\%$ fidelity, and the Hidden Shift circuit with up to 4 qubits with $>60\%$ fidelity. Some accuracy was retained even with larger circuit widths, with a simulated 11-qubit implementation of the Bernstein-Vazirani and Deustch-Jozsa algorithms giving $>50\%$ fidelity. Since the hidden-shift algorithm presented in the benchmark entangles only 2 qubits at a time, there is no difference between the performance of all-to-all versus nearest-neighbor connectivities in this circuit. However, all-to-all connectivity significantly increased the fidelity of wider circuits in the Deutsch-Jozsa and Bernstein-Vazirani algorithms. The most significant increases in fidelity occurred within the 7-11 qubit range, where the circuit fidelity increases on average by $2.6\%$ ($4.3\%$ max) on the Deutsch-Jozsa circuits and $3.1\%$ (3.7\% max) on the Bernstein-Vazirani circuits.

 \begin{figure*}
    \centering
    \includegraphics[width=16.cm]{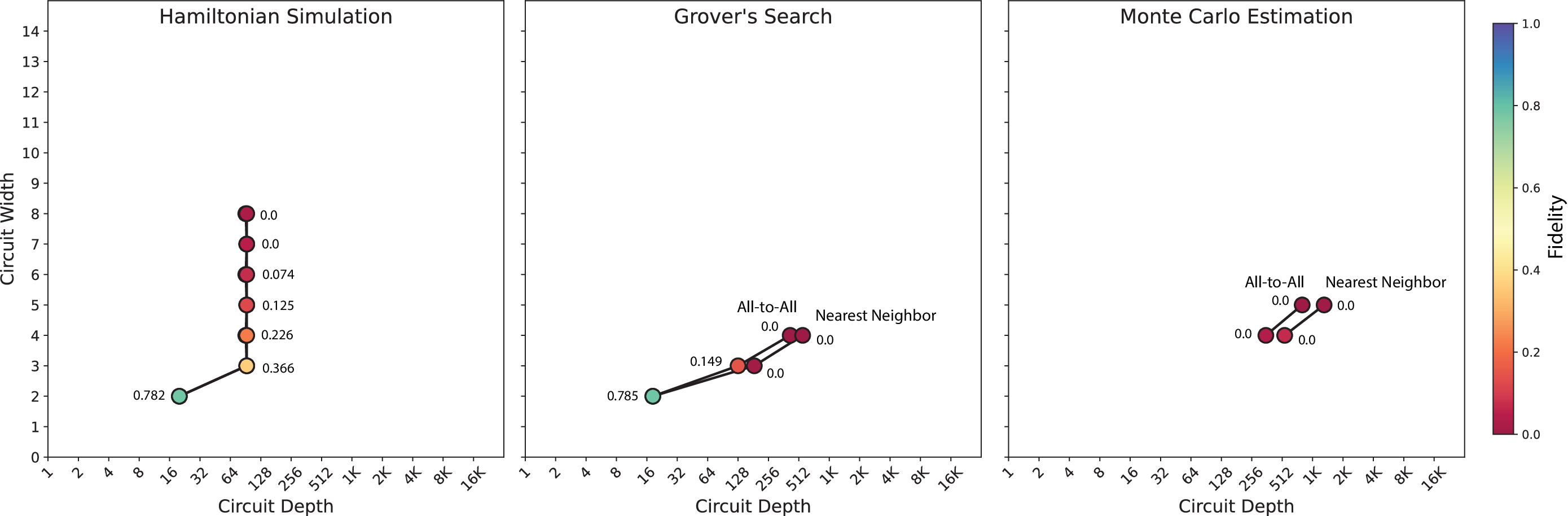}
    \caption{Simulation results for  Hamiltonian simulation, Grover's search,  and Monte Carlo estimation. For Hamiltonian simulation there was negligible difference between all-to-all and nearest-neighbor circuits. 
    }
    \label{fig:application}
\end{figure*}

The results of quantum subroutines are shown in Fig. \ref{fig:subroutines}. With only nearest-neighbor connectivity, Method 2 of the quantum Fourier transform benchmark reached a maximum of $85\%$ fidelity with 2 qubits falling to $67\%$ fidelity with 3 qubits. The simulation achieved an $89\%$ fidelity with a 3-qubit phase estimation circuit. Method 1 of the QFT benchmark reached a maximum fidelity of $77$\% with 2 qubits falling to a $48$\% fidelity with 3 qubits. All amplitude estimation circuits produced results with fidelity $<20\%$. All-to-all connectivity also significantly improved the performance of quantum subroutines. The most significant increases in fidelity were found for Method 2 of the quantum Fourier transform benchmark and phase estimation. On average, the fidelity of the inverse quantum Fourier transform (Method 2) increased by $13\%$ in the 3-5 qubit range, peaking at an increase of $17\%$ at the 5-qubit implementation. The average increase in fidelity of phase estimation circuits on the same range was $22\%$, and peaks at a $26\%$ increase at the 4 qubit implementation.

The results for several applications are shown in Fig. \ref{fig:application}. Both the Grover's search and Hamiltonian estimation had high-fidelity results for 2 qubits, with both Grover's search and the Hamiltonian simulation giving $78\%$ fidelity on average. Implementations of the Hamiltonian simulation with 3-4 qubits are lower than that of the 2-qubit implementation, with the maximum fidelity being $36$\% in the 3-qubit implementation. All Hamiltonian circuits with 5 or more qubits, along with all other circuits with 3 or more qubits, returned an average fidelity of $<15\%$ for both nearest-neighbor and all-to-all connectivity simulations. The nearest neighbor circuits were identical for all Hamiltonian circuits and the $n=2$ Grover's search circuit, and all other results had too low a fidelity to be meaningfully compared between all-to-all and nearest neighbor implementations.

\section{Conclusion} 
\label{sec:conc}

We have presented an error model which is consistent with experimental data and which can be used to simulate circuits with arbitrary widths and depths. When simulating the processor under different connectivity constraints, we found significant improvement in circuit fidelities as the connectivity of the computer increases. This improvement is consistent throughout several types of quantum circuits included in the QED-C benchmark, particularly with quantum subroutines and oracle-based algorithms.

Our inclusion of a proposed noise model provides an additional multifaceted, quantitative assessment of our computer's performance. It allows for us to determine specific sources of noise which are more prevalent than others, easily simulate how changes in one parameter may affect many different types of circuits, and instantly project the results of proposed experiments. In this way, using this noise model as a preliminary check in experimental work helps expedite the process of designing and improving our quantum processor.

Additionally, the benefits of a different proposed qubit topology than a square grid can be assessed without having to fundamentally change the computer. This could include studying and quantifying the benefits of a larger projected radius over which Rydberg blockage would occur or different layouts such as a triangular optical lattice. As developing a scalable quantum processor becomes more pertinent, the benefits and drawbacks of the qubit topology become more significant, making studies such as these more important to carry out.

We plan on continuing to study quantum subroutines on the computer, since they showed the most significant improvements. For example, the SupermarQ benchmark \cite{Tomesh2022}, provides a set of mid-width circuits not used in the QED-C benchmark that would be interesting to run under our noise model.

\acknowledgements

This material is based upon work supported by    NSF Award 2016136 for the QLCI center Hybrid Quantum Architectures and Networks, NSF award No. 2210437, and the U.S. Department of Energy Office of Science National Quantum Information Science Research Centers.


\bibliography{qc_refs.bib,rydberg.bib,saffman_refs.bib}

\appendix

\section{Noise Model}
\label{ssec:noisemodel}
One can simulate a $n$-ququart lossy channel using a set of at most $16^n$ Kraus operators $\{ {\sf A}_i \}$. The channel transforms the density matrix of the system as
\begin{equation}\rho \rightarrow \sum_{i} {\sf A}_i\rho {\sf A}_i^{\dag}\end{equation}
where $\sum_{i} {\sf A}_i^{\dag}{\sf A}_i = \sf I$.
The model uses 6 distinct channels in total, with both SPAM channels using the bit-flip error.

Throughout the appendix, we will generalize Pauli operators in the extended basis with a loss state $\{\ket{0},\ket{1},\ket{l}_0,\ket{l}_1  \}$ as 
\begin{equation}
    {\sf X} = \begin{bmatrix}
    0 & 1 & 0 & 0\\
    1 & 0 & 0 & 0\\
    0 & 0 & 1 & 0\\
    0 & 0 & 0 & 1
    \end{bmatrix}, {\sf Y} = \begin{bmatrix}
    0 & -i & 0 & 0\\
    i & 0 & 0 & 0\\
    0 & 0 & 1 & 0\\
    0 & 0 & 0 & 1
    \end{bmatrix}, {\sf Z} = \begin{bmatrix}
    1 & 0 & 0 & 0\\
    0 & -1 & 0 & 0\\
    0 & 0 & 1 & 0\\
    0 & 0 & 0 & 1
    \end{bmatrix}
\end{equation}
In general, the error probability $p$ of a gate with an error rate $r$ per $\pi$ pulse, such as the $\sf{R}_\phi$ and $\sf{R}_Z$ gates, is
\begin{equation}
p=r\frac{\theta}{\pi}
\end{equation}
Parameters for error probabilities, and for $T_1, T_2^\ast$ times, are listed in Table \ref{tab:table1}.

\subsection{Depolarization error}
A depolarization channel with error probability $p$ acts upon a density matrix as
\begin{equation}
    \rho \rightarrow (1-p)\rho + \frac{p}{3}{\sf X}\rho {\sf X} + \frac{p}{3} {\sf Y} \rho {\sf Y} + \frac{p}{3} {\sf Z} \rho {\sf Z}
\end{equation}
This is the only type of error present on the microwave gates, and since the error rate is low, this type of error does not contribute significantly to the benchmark results. 

\subsection{Phase-flip error}
A phase-flip channel with error probability $p$ acts upon a density matrix as
\begin{equation}
    \rho \rightarrow (1-p)\rho + p {\sf Z}\rho {\sf Z}
\end{equation}
This noise channel is present on both the local ${\sf R}_z$ gate and the Rydberg gate.

\subsection{Qubit loss error}
The action of a qubit loss channel with error probability $p$ on a density matrix can be represented as
\begin{equation}
\rho \rightarrow \sf{A}_0\rho \sf{A}_0 + \sf{A}_1\rho \sf{A}_1^\dagger
\end{equation}
where
\begin{equation}
\sf{A}_0=\begin{bmatrix}
1 & 0 & 0 & 0\\
0 & \sqrt{1-p} & 0 & 0\\
0 & 0 & 1 & 0\\
0 & 0 & 0 & 1
\end{bmatrix}
\end{equation}
and
\begin{equation},\sf{A}_0=\begin{bmatrix}
0 & 0 & 0 & 0\\
0 & 0 & 0 & 0\\
0 & \sqrt{p} & 0 & 0\\
0 & 0 & 0 & 0
\end{bmatrix}
\end{equation}
for a loss to dark state channel and
\begin{equation},\sf{A}_0=\begin{bmatrix}
0 & 0 & 0 & 0\\
0 & 0 & 0 & 0\\
0 & 0 & 0 & 0\\
0 & \sqrt{p} & 0 & 0
\end{bmatrix}
\end{equation}
for a loss to bright state channel.
\subsection{Qubit decay}

Qubit decay decoheres qubits in the $\ket{1}$ basis to the $\ket{0}$ basis. A qubit decay channel with error rate $p$ acts upon the density matrix as
\begin{equation}
    \rho \rightarrow {\sf A}_0\rho {\sf A}_0 + {\sf A}_1 \rho {\sf A}_1
\end{equation}
with
\begin{equation}
    {\sf A}_0 = \begin{bmatrix}
    1 & 0 & 0 & 0\\
    0 & \sqrt{1-p} & 0 & 0\\
    0 & 0 & 1 & 0\\
    0 & 0 & 0 & 1
    \end{bmatrix},{\sf A}_1=\begin{bmatrix}
    0 & \sqrt{p} & 0 & 0\\
    0 & 0 & 0 & 0\\
    0 & 0 & 0 & 0\\
    0 & 0 & 0 & 0
    \end{bmatrix}
\end{equation}

\subsection{Bit-flip}
A bit-flip error operation is used to simulate  qubit preparation error. The channel's action on the qubit can be given by
\begin{equation}
    \rho \rightarrow (1-p)\rho + p {\sf X}\rho {\sf X}
\end{equation}
Qubit measurement error is performed after the bitstring populations are collected from the density matrix, reducing computation time. The error channel simply flips a bit in the measured bitstring with probability $p$ for each bit.

\subsection{Decoherence}
Given a lossy density matrix
\begin{equation}
    \rho = \begin{bmatrix}
    \rho_{00} & \rho_{01}\\
    \rho_{10} & \rho_{11}
    \end{bmatrix}
\end{equation}
one can describe the action of the $T_1, T_2^\ast$  decoherence channels directly on the matrix as
\begin{equation}
    \rho \rightarrow \begin{bmatrix}
    d_1\rho_{00} + (1-d_1)p_0p_t & d_2\rho_{01}\\
    d_2\rho_{10} & d_1\rho_{11}+(1-d_1)p_1p_t
    \end{bmatrix}
\end{equation}
with 
\begin{gather}
    d_1 = e^{-\frac{t}{T_1}}\\
    d_2 = e^{-\frac{t}{T_2^\ast}}\\
    p_1 = 1 - p_0
\end{gather}
and
\begin{equation}
    p_t = \rho_{00}+\rho_{11}
\end{equation}
being the total population in the $\ket{0}$ and $\ket{1}$ states, where $p_0$ is the equilibrium population of $\ket{0}$ for the $T_1$ channel. To simplify the complexity of this channel for numerical computation, we describe it as a composition of two Kraus channels:
\begin{equation}\begin{aligned} \rho \rightarrow {\sf A}\rho {\sf A}^{\dag} + {\sf B} \rho {\sf B}^{\dag} + {\sf C}\rho {\sf C}^{\dag}\\
\rightarrow (1-\phi)\rho + \phi {\sf Z}\rho {\sf Z}\end{aligned}\end{equation}
with
\begin{equation}
    {\sf A} = \begin{bmatrix} \sqrt{p_0(1-d_1) + d_1} & 0 & 0 & 0\\
    0 & \sqrt{p_1(1-d_1) + d_1} & 0 & 0\\
    0 & 0 & 1 & 0\\ 0 & 0 & 0 & 1\end{bmatrix}
\end{equation}
\begin{equation}
    {\sf B} = \begin{bmatrix} 0 & \sqrt{p_0(1-d_1)} & 0 & 0\\
    0 & 0 & 0 & 0\\
    0 & 0 & 0 & 0\\
    0 & 0 & 0 & 0\end{bmatrix}
\end{equation}
\begin{equation}
    {\sf C} = \begin{bmatrix} 0 & 0 & 0 & 0\\
    \sqrt{p_1(1-d_1)} & 0 & 0 & 0\\
    0 & 0 & 0 & 0\\
    0 & 0 & 0 & 0\end{bmatrix}
\end{equation}
\begin{equation}
    \phi = \frac{1}{2} - \frac{d_2}{2 \sqrt{(p_0(1-d_1) + d_1) (p_1(1-d_1) + d_1)}}
\end{equation}
Note that, since the only mechanism for a qubit to enter the $\ket{l}_0$ or $\ket{l}_1$ state is the loss operator, all qubits in the $\ket{l}_0$ or $\ket{l}_1$ states will be incoherent with states where this qubit is in the $\ket{0}$ or $\ket{1}$ state. Therefore, instead of $16^n$ nonzero entries in the density matrix, there are only $6^n$ for each density matrix. This means that the circuits can be simulated much quicker than usual ququart implementations, since some channels require fewer Kraus operators.

The noise model we use always applies the gate itself first, and the $T_1, T_2$  channels last. For the microwave and local ${\sf R}_z$ gates, only one error is applied between the two operations. These errors are depolarization error and a phase-flip error respectively. The $\sf C_Z$ gate applies the qubit loss channel directly after the gate, followed by the bit decay channel, the phase-flip channel, and finally the correlated phase-flip channel. 

\section{Methodology}

Results were obtained by simulating the density matrix evolution of each circuit, with  the Kraus operators described in Appendix \ref{ssec:noisemodel}. Noise parameters were determined by maximizing the value of the average classical fidelities between the simulated and experimental results of the three QAOA circuits in \cite{Graham2022}, and are listed in Table \ref{tab:table1}. We chose the classical fidelity as our performance metric as it is directly connected to the metric used by the QED-C benchmark to test circuit accuracy.

\begin{figure*}[!t]
    \centering
    \includegraphics[scale=0.35]{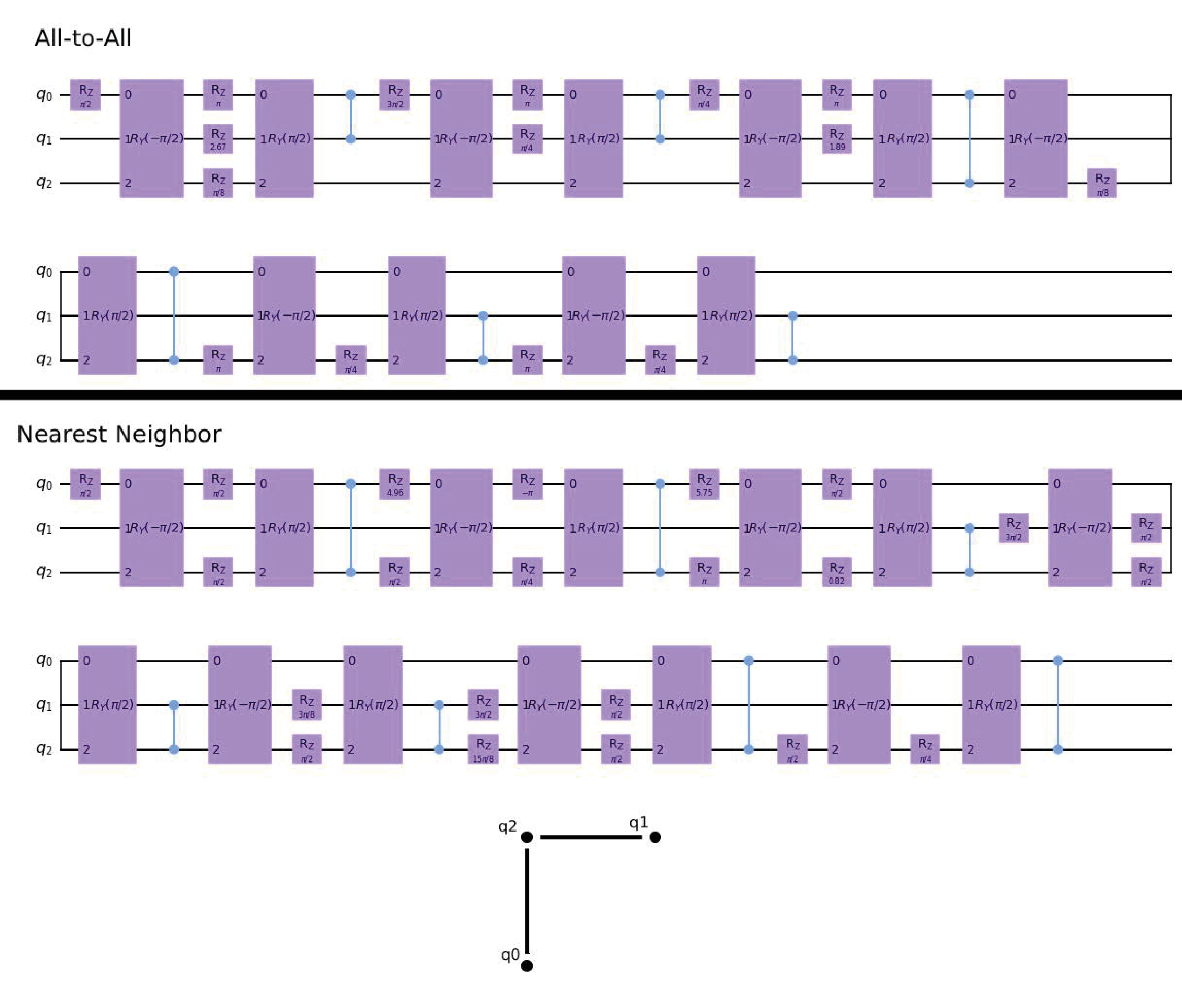}
    \caption{Sample circuit of a 3-qubit inverse quantum Fourier transform. Both the all-to-all and nearest-neighbor circuits are shown, with the nearest-neighbor topology shown below the circuit. The operations covering three qubits in one time slice are global microwave rotation gates ${\sf R}_\phi^G(\theta).$  }
    \label{fig:qft}
\end{figure*}

\begin{figure*}[!t]
    \centering
    \includegraphics[scale=0.35]{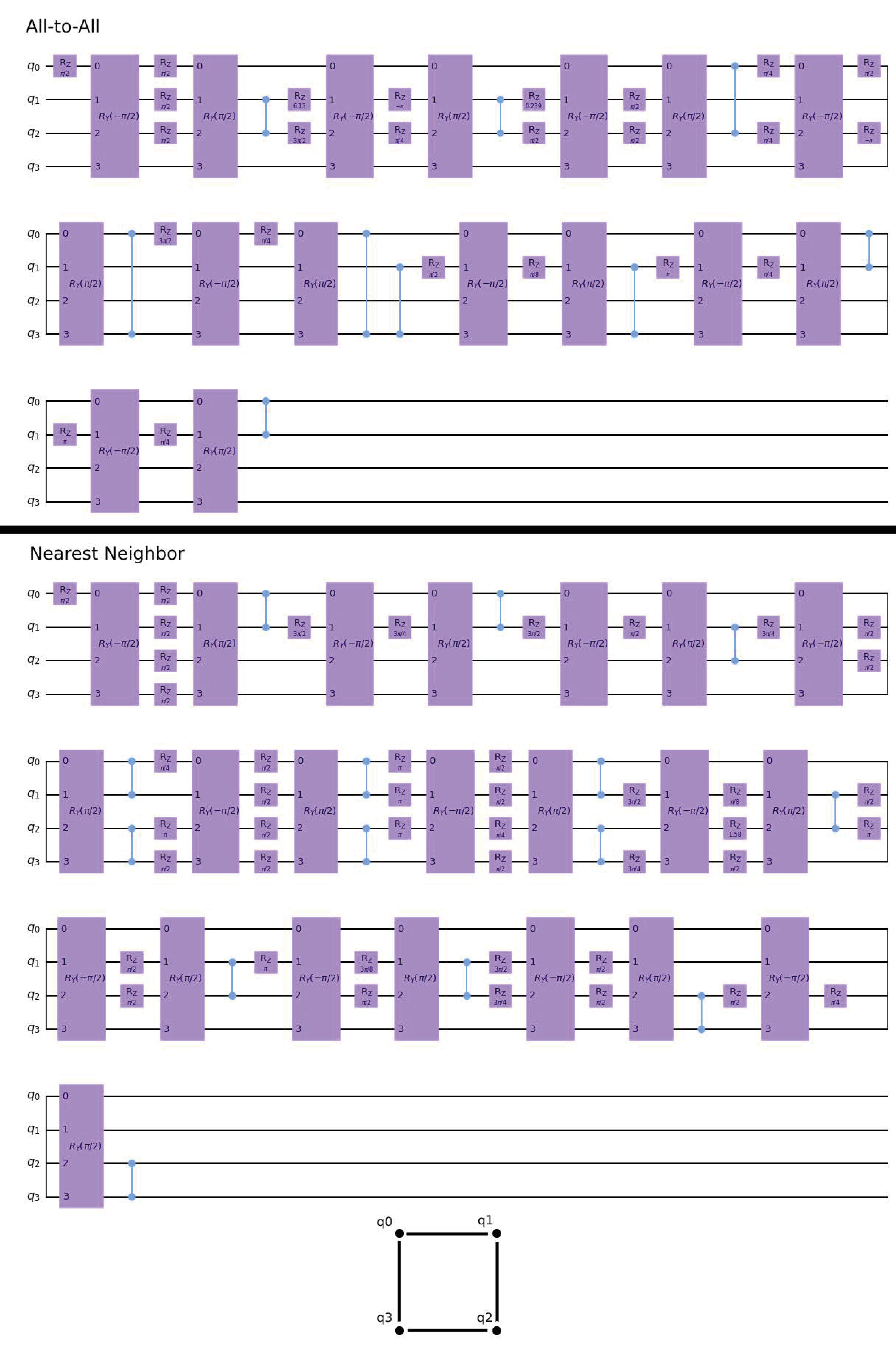}
    \caption{Sample circuit of a 4-qubit phase estimation circuit. Both the all-to-all and nearest-neighbor circuits are shown, with the nearest-neighbor topology shown below the circuit.}
    \label{fig:phaseest}
\end{figure*}

The structure of the circuits provided in the benchmark depend on a variable parameter, meaning that two circuits of the same type and width can have different fidelities. Specific examples of this parameter include the phase to be measured in the phase estimation circuits, and the bitstring used in the Deutsch-Jozsa and Bernstein-Vazirani circuits. Because of this, for circuits that allowed 3 or more values for the parameter, we sampled 3 circuits per data point, corresponding to 3 distinct values chosen at random. We then took the resultant fidelity as listed in Figures \ref{fig:oracle}, \ref{fig:subroutines}, \ref{fig:application} as the average of the fidelities of the three runs. For circuits that admitted less than 3 values, we instead ran one circuit for each value. Circuits were sampled using the source code provided in \cite{Lubinski2021}. Exceptions to this triple sampling are the amplitude estimation circuits, where only two circuits per data point were chosen, and Monte Carlo circuits, where only one circuit per data point was chosen to reduce processing time. These circuits had low fidelity in our simulations so further sampling was not required. Sample circuits are given in Figs. \ref{fig:qft} and \ref{fig:phaseest}. 


As the $T_1$  and $T_2$  times were measured precisely, both the $T_1$ and $T_2$ times were fixed, while the rest of the noise model parameters could vary. The goodness of fit between the two models was quantified by the classical fidelity between both distributions, which is closely related to the metric used in the benchmark. We optimized the parameters using the Nelder-Mead algorithm. As the simulator determines the exact resultant density matrix of the circuit, the populations of simulated states could be read directly after one evaluation of the circuit.

On average, the classical fidelity between the simulated and measured results of the three MAXCUT circuits is 98.7\%, with the lowest being 97.5\% on the $p=3$ circuit.  The fidelities of the simulated GHZ states overestimated the experimental fidelities by 8.0\% and differed on average from the fidelities of the measured GHZ states by 
about 5.5\%.

\newpage

\section{Treating qubit loss}
\label{app.C}

The result of multiple runs of a benchmark circuit is a set of probabilities $P_{\rm out}(x)$ for the observation of bit string $x$,  with $\sum_x P_{\rm out}(x)=1.$ Note that the fidelity used as a figure of merit in the QED-C benchmark only examines the probability distribution of reported bitstrings for a given circuit, not the final quantum state of the system. This is beneficial as it enables real hardware to be evaluated using the benchmark solely by repeatedly running circuits to acquire the probability distribution rather than using tomography or other similarly expensive means to reconstruct the final quantum state, but does introduce an insensitivity to certain types of errors. For example, any error in the relative phase of the computational basis states is masked completely.

 Atoms that are lost from their traps during circuit execution will be interpreted by the atom readout procedure used in \cite{Graham2022} as being in the $\ket{1}$ state. This is the fate of the majority of atoms captured by our catch-all bright lost state $\ket{l}_1$ and dark lost state $\ket{l}_0$. Thus for both calibrating the error model and when calculating the fidelity of circuits in the benchmark, we reduce the ququart probability distribution to a qubit probability distribution by replacing the $\ket{l}_1$ state with the $\ket{1}$ state and the $\ket{l}_0$ state with the $\ket{0}$ state to mimic what we would receive as output from real hardware.

\end{document}